\documentclass{osa-article}

\journal{osajournal}

\articletype{Research Article}

\usepackage{graphicx}
\usepackage{amsmath,amssymb,amsfonts}

\usepackage{lineno}

\newcommand{\del}{\partial}

\newcommand{\bs}{\boldsymbol}
\newcommand{\rmd}{\mathrm{d}}
\newcommand{\rmi}{\mathrm{i}}
\newcommand{\rminc}{\mathrm{inc}}
\newcommand{\rmsc}{\mathrm{sc}}
\newcommand{\rmtot}{\mathrm{tot}}

\begin{document}


\title{Robust Field-Only Surface Integral Equations: {S}cattering from a Perfect Electric Conductor}   

\author{Qiang Sun, \authormark{1,2,*}
  Evert Klaseboer, \authormark{3}
  Alex J.~Yuffa, \authormark{4}
  and Derek Y.~C.~Chan, \authormark{5,6}}

\address{\authormark{1} Centre of Excellence for Nanoscale BioPhotonics, RMIT University, Melbourne, VIC 3001, Australia\\
  \authormark{2} Department of Chemical Engineering, The University of Melbourne, Parkville 3010, VIC, Australia\\
  \authormark{3}  Institute of High Performance Computing, 1 Fusionopolis Way, Singapore 138632, Singapore\\
  \authormark{4} National Institute of Standards and Technology, Boulder, CO 80305 USA\\
  \authormark{5} School of Mathematics and Statistics, The University of Melbourne, Parkville 3010, VIC, Australia\\
  \authormark{6} Department of Mathematics, Swinburne University of Technology, Hawthorn VIC 3121 Australia
}

\email{\authormark{*}qiang.sun@rmit.edu.au}



\begin{abstract*}
A robust field-only boundary integral formulation of electromagnetics is derived without the use of surface currents that appear in the Stratton--Chu formulation. For scattering by a perfect electrical conductor (PEC), the components of the electric field are obtained directly from surface integral equation solutions of three scalar Helmholtz equations for the field components. The divergence-free condition is enforced via a boundary condition on the normal component of the field and its normal derivative. Field values and their normal derivatives at the surface of the PEC are obtained directly from surface integral equations that do not contain divergent kernels. Consequently, high-order elements with fewer degrees of freedom can be used to represent surface features to a higher precision than the traditional planar elements.  This theoretical framework is illustrated with numerical examples that provide further physical insight into the role of the surface curvature in scattering problems.  
\end{abstract*}


\section{Introduction}\label{sec:introduction} 
The macroscopic Maxwell equations describe the interaction between electromagnetic waves and matter that impacts on almost every aspect of our daily life ranging from daylight~\cite{Rayleigh1881,Rayleigh1899} to  communication~\cite{Sarkar2018}, food processing~\cite{Chandrasekaran2011}, energy harvesting~\cite{Hou2012,Jang2016,Yu2017} as well as biomedical sensing and therapy~\cite{Homola2008,Mayer2011,Baumann2016}.  Because it is only possible to obtain analytical solutions to Maxwell's equations for scatterers with canonical geometries (e.g., planes, cylinders, and spheres~\cite{Mie1908,Bohren1983,FrezzaJOSA2018}), an efficient numerical solution is required to tackle practical problems. In general, numerical solutions to Maxwell's equations in linear, homogeneous media fall into two categories: methods that discretize the 3D spatial domain such as the finite difference time domain method~\cite{Yee1966,Taflove1988} and the finite element method~\cite{Jin2014}, or methods based on 2D surface integral formulation methods using Green's theorem~\cite{Gibson2008}. When compared to 3D domain methods, the surface integral methods have some distinct advantages. The most obvious advantage is that the number of dimensions in a problem is reduced by one. Another advantage is that for exterior problems, the Sommerfeld radiation condition~\cite{Schot1992} is satisfied directly without the need to use artificial absorbing boundary conditions at infinity.
  
In general, the current popular industry-standard surface integral approaches are based on the Stratton--Chu~\cite{Stratton1939,Stratton1941} or the PMCHWT~\cite{Poggio1973,Wu1977,Chang1977,uysalJOSA2016} formulation that involve solving for the surface electrical current density $\bs{J}$ and the magnetic current density $\bs{M}$~\cite{NiuJOSAA2014, CuiJOSAA2014, ForestiereJOSAA2012, Kern2009} as the intermediate boundary unknowns. In these approaches, the divergence-free condition on the fields is satisfied implicitly. To enforce current conservation at each surface element, the RWG vector basis functions~\cite{Rao1982} are commonly used to represent the current densities.  Due to the use of the continuity equation between the surface charge and the surface current, such formulations suffer from numerical instabilities in the long wavelength limit that is also referred to as the zero frequency catastrophe~\cite{Vico2016, Zhao2000}. The surface integral equations have integrands with strong singularities~\cite{Chew1989} that preclude the accurate evaluation of the field or its derivatives at or near the surface of the scatterer.

Another formulation of the surface integral method is based on the scalar and vector potentials $\phi$ and $\boldsymbol{A}$~\cite{GarciadeAbajo2002, Vico2016}, where the surface charge and the surface current are chosen as the boundary unknowns. Although these integral equations still involve singular kernels, they do not suffer from numerical instabilities at or near zero frequency. Therefore, the $\phi$ and $\boldsymbol{A}$ formulation is often used in plasmonic applications where the size of the scatterer is small relative to the incident wavelength.

In a physical scattering problem, the field quantities are finite and well-defined at material boundaries. Thus, it seems reasonable that it should be possible to develop a surface integral formulation that does not contain singularities in the integrals; after all, the origin of the singularities is mathematical and does not have a physical basis. Furthermore, in applications such as micro-photonics it is desirable to be able to obtain directly accurate values of the electromagnetic fields and their derivatives on boundaries without further post-processing.


Recently, there have been two independent and successful attempts to develop a boundary integral method that solves Maxwell's equations \emph{directly} for the electric field, $\bs{E}$. These approaches are very different from the surface integral methods reviewed above. In one approach, the divergence-free condition on the electric field is replaced by a scalar Helmholtz equation for $\bs{r} \cdot \boldsymbol{E}$ via a vector identity~\cite{Klaseboer2017, Sun2017}, where $\bs{r}$ is a space position vector. This, together with the scalar Helmholtz equation for each of the three Cartesian components of $\bs{E}$, provides a set of \emph{four} coupled scalar wave equations to be solved.  In the other approach, the divergence-free condition is replaced with an equivalent boundary condition~\cite{Yuffa2006, Yuffa2018, Yuffa2019, Yuffa2019a}.  This formulation directly leads to \emph{three} scalar integral equations.  However, the kernels of these three integral equations are singular.

In this paper, we develop a fully desingularized surface integral formulation for electromagnetic scattering that facilitates the direct solution of the electric or magnetic field and their normal derivatives at the material boundaries.  This is accomplished by introducing the fully desingularized boundary integral method~\cite{Klaseboer2017, Sun2017} into the singular integral formulation based on the divergence-free boundary condition~\cite{Yuffa2006, Yuffa2018, Yuffa2019, Yuffa2019a}.  In other words, we combine the two methods mentioned in the previous paragraph to obtain a boundary integral method containing \emph{three} coupled scalar \emph{nonsingular} surface integral equations. This formulation is consistent with the notion that given the scattering problem entails no singular behavior at boundaries, the theoretical formulation should not contain mathematical singularities either.

The physical basis of this formulation is conceptually simple and can provide direct access to values of the field and its normal derivative at boundaries. The absence of mathematical singularities allows for the use of simple algorithms that are efficient and accurate. In order not to obscure the physical and mathematical simplicity of the method, we only consider smooth perfect electrical conductor (PEC) scatterers.  The extension of this method to dielectric scatterers will be provided in our forthcoming paper~\cite{SunFuture}.

The paper is organized as follows.  In Section~\ref{sec:formulation}, we present the mathematical formulation of the method together with a simple physical interpretation of the key boundary condition.  In Section~\ref{sec:comp-advant-appl}, we discuss the numerical formulation of the method and, in Section~\ref{sec:illustr-numer-result}, we present some numerical examples.  Finally, in Section~\ref{sec:conclusions}, we summarize the main features of the method and conclude the paper.

\section{Formulation}\label{sec:formulation}
We illustrate our formulation with the scattering of an incident plane wave by 3D perfect electrical conductors. In the frequency domain with time dependence $\exp(-\rmi \omega t)$, the propagating scattered electric field $\bs{E}^\rmsc$ in a source-free region is given by
\begin{subequations}
  \label{eq:Helmholtz_E}
  \begin{equation}
    \label{eq:Helmholtz_E_part_a}
    \nabla^2 \bs{E}^\rmsc + k^2 \bs{E}^\rmsc = \bs{0},
  \end{equation}
  \begin{equation}
    \label{eq:Helmholtz_E_part_b}
    \bs{\nabla} \cdot \bs{E}^\rmsc = 0,
  \end{equation}
\end{subequations}
where the wavenumber $k = \sqrt{\epsilon \mu} \omega$ with $\epsilon$ and $\mu$ being, respectively, the permittivity and permeability of the medium.  The incident field $\bs{E}^\rminc$ and the total field $\bs{E}^{\text{tot}}=\bs{E}^{\text{inc}}+\bs{E}^\rmsc$ also satisfy \eqref{eq:Helmholtz_E}.  The divergence-free condition \eqref{eq:Helmholtz_E_part_b} implies there are only two independent components of $\bs{E}^\rmsc$. Given an incident wave, such as a plane wave $\bs{E}^{\text{inc}} = \bs{E}_0 \exp(\rmi \bs{k} \cdot \bs{r})$, (where $\bs{r} = (x, y, z)$ is the position vector, $\bs E_0 \cdot \bs k=0$, $|\bs k|= k$), \eqref{eq:Helmholtz_E} can be solved by imposing the boundary conditions.  These boundary conditions require the two independent tangential components of $\bs{E}^\rmtot$ to vanish on the surface of the PEC and the scattered field, $\bs{E}^\rmsc$, to satisfy the Silver--M\"{u}ller radiation condition~\cite{Colton1992} 
at infinity.

The solution of the vector Helmholtz equation \eqref{eq:Helmholtz_E_part_a} can be expressed as the solution of the surface integral equation for each Cartesian component of the scattered field, i.e., $\{E^{\rmsc}_{x}, E^{\rmsc}_{y}, E^{\rmsc}_{z}\}$.  To see this, we apply Green's second identity outside the PEC scatterer to obtain
\begin{align} \label{eq:BIE_Ei}
    c_0(\bs{r}_0) E^{\rmsc}_\alpha(\bs{r}_0) &+ \int_S E^{\rmsc}_\alpha(\bs{r}) \frac{\del G(\bs{r}, \bs{r}_0)}{\del n} \, \rmd S(\bs{r}) = \int_S  \frac{\del E^{\rmsc}_\alpha(\bs{r})}{\del n} G(\bs{r}, \bs{r}_0) \, \rmd S(\bs{r}), \quad \alpha=x,y,z,
\end{align}
where $G(\bs{r}, \bs{r}_0) = \exp(\rmi k|\bs{r} - \bs{r}_0| )/|\bs{r} - \bs{r}_0|$ is the free-space Green's function, $\del/\del n = \bs{n} \cdot \nabla$ is the normal derivative, and $\bs{n}$ is the unit normal that points into the scatterer. The integration (source) point $\bs{r}$ is on the surface of the scatterer and if $\bs{r}_0$ is located outside the scatterer, then $c_0 = 4\pi$.  If $r_0$ is on the surface $S$, then $c_0$ is the solid angle subtended at $\bs{r}_0$ and is equal to $2\pi$ if the tangent plane is continuous at $\bs{r}_0$.

It is important to realize that a solution to \eqref{eq:BIE_Ei} may or may not satisfy the divergence-free condition \eqref{eq:Helmholtz_E_part_b}.  An elegant way to ensure that the electric field satisfies the divergence-free condition in a 3D domain was proposed by Yuffa and Markkanen~\cite{Yuffa2018}. It uses the fact that if the divergence-free condition exists at all points on the surface $S$, then the electric field is divergence-free everywhere in the 3D domain.  This fact was rigorously proved in their paper~\cite{Yuffa2018}; however, their proof is rooted in differential geometry and is rather technical. Thus, we offer an alternative proof based directly on the fact that $\bs{\nabla} \cdot \bs{E}^\rmsc$ satisfies the scalar Helmholtz equation. To see this, we note that $\bs{E}^\rmsc$ is differentiable and satisfies the vector Helmholtz equation \eqref{eq:Helmholtz_E_part_a}. Thus, any partial derivative of $\bs{E}^\rmsc$ will also satisfy \eqref{eq:Helmholtz_E_part_a}, and therefore we can write 
\begin{equation}
  \label{eq:divE_wave}
  \nabla^2 (\bs{\nabla} \cdot \bs{E}^\rmsc)
  + k^2 ( \bs{\nabla} \cdot \bs{E}^\rmsc) = 0.    
\end{equation}
The solution to \eqref{eq:divE_wave} can be expressed in the usual surface integral representation via Green's second identity to obtain
\begin{equation}
  \label{eq:BIE_divV} 
  c_0(\bs{r}_0) \left[\bs{\nabla} \cdot  \bs{E}^\rmsc(\bs{r}_0)\right]
  + \int_S \left[\bs{\nabla} \cdot \bs{E}^\rmsc(\bs{r})\right]
  \frac{\partial G(\bs{r}, \bs{r}_0)}{\partial n} \, \rmd S(\bs{r})
  = \int_S
  \frac{\partial \left[\bs{\nabla} \cdot \bs{E}^\rmsc(\bs{r})\right]}{\partial n}
  G(\bs{r}, \bs{r}_0) \, \rmd S(\bs{r}).
\end{equation}
The above equation expresses the value of $\bs{\nabla} \cdot \bs{E}^\rmsc$ at an arbitrary point $\bs{r}_0$ outside of the scatterer in terms of the values of $\bs{\nabla} \cdot \bs{E}^\rmsc$ and its normal derivative $\partial [\bs{\nabla} \cdot \bs{E}^\rmsc]/\partial n$ on the surface $S$ ($\bs{r}$ is always located on $S$).  From \eqref{eq:BIE_divV}, we see that if $\bs{\nabla} \cdot \bs{E}^\rmsc$ vanishes on $S$ and we set $\bs{r}_0$ on $S$, then the left hand side of \eqref{eq:BIE_divV} is zero. Therefore, $\partial [\bs{\nabla} \cdot \bs{E}^\rmsc]/\partial n$ is zero as well on the surface $S$ provided that the frequency is different from any of the internal resonance frequencies \cite{Peterson1990,Colton1992}. Now, applying \eqref{eq:BIE_divV} with $\bs{r}_0$ outside of the scatterer and not on the surface $S$, we conclude that $\bs{\nabla} \cdot \bs{E}^\rmsc(\bs{r}_0) = 0$ outside of the scatterer provided that $\bs{\nabla} \cdot \bs{E}^\rmsc$ vanishes on $S$. 

The value of $\bs{\nabla} \cdot \bs{E}^\rmtot$ on the surface, or rather its limiting value as one approaches the surface, may be found by first decomposing $\bs{E}^{\text{tot}}$ along the normal $\bs{n}$ and the two tangential unit vectors, $\{\bs{t}_1,\bs{t}_2\}$, and then computing the divergence.  The details of this derivation are provided in \ref{Appendix_zero_div} with the final result given by \eqref{eq:A_final_result} (also see (23) in \cite{Yuffa2018}), i.e.,  
\begin{equation}
  \label{eq:1}
  \bs{n} \cdot \frac{\partial { \bs{E}^\rmtot }  }{ \partial{n} }
  - \kappa E^\rmtot_{n} + \frac{\partial { E^\rmtot_{t_{1}}  }} { \partial{t_{1}} } + \frac{\partial { E^\rmtot_{t_{2}}  }} { \partial{t_{2}} }=0,
\end{equation}
where $\kappa$ is the mean curvature.  In \eqref{eq:1}, $E^\rmtot_n = \bs{n} \cdot \bs{E}^\rmtot$ and $\{E^\rmtot_{{t}_1} = \bs{t}_1 \cdot \bs{E}^\rmtot, E^\rmtot_{t_2} = \bs{t}_2 \cdot \bs{E}^\rmtot \}$ are the normal and tangential components of $\bs{E}^\rmtot$, respectively, and $\{\partial /\partial t_1, \partial /\partial t_2\}$ are the tangential derivatives on the surface.  On the surface of the PEC scatterer, the tangential components of $\bs{E}^\rmtot$ vanish and \eqref{eq:1} reduces to
\begin{equation}
  \label{eq:E_dE/dn_BC}
  \bs{n} \cdot \frac{\partial { \bs{E}^\rmtot }  }{ \partial{n} } =
  \kappa E^\rmtot_n.
\end{equation}
From \eqref{eq:E_dE/dn_BC}, we see that on the surface of the PEC scatterer the normal component of the normal derivative of $\bs{E}^\rmtot$ and the normal component of $\bs{E}^\rmtot$ are proportional to each other.  This is a direct consequence of the boundary condition $E^\rmtot_{t_1}=E^\rmtot_{t_2}=0$ and the divergence-free condition, $\bs{\nabla} \cdot \bs{E}^\rmtot =0$ on $S$.  This relation will also ensure that $\bs{E}^\rmtot$ is divergence-free in the 3D domain outside the scatterer.

The result \eqref{eq:E_dE/dn_BC} has a simple physical interpretation. From elementary electrostatics, the field emanating from a charged \emph{infinite planar} PEC is constant, i.e., $\del \bs{E}^\rmtot/\del n = \bs{0}$, and is directed normal to the surface. In \eqref{eq:E_dE/dn_BC}, $E^\rmtot_n $ is the induced charged density on the PEC surface and so $ \bs{n} \cdot \del \bs{E}^\rmtot/\del n$ can only be non-zero if the PEC has a non-zero curvature $\kappa$.

In summary, the electric field due to scattering by PEC scatterers can be found by solving the three scalar surface integral equations contained in \eqref{eq:BIE_Ei} for each of the Cartesian components of the scattered field $\bs{E}^\rmsc$ and imposing the boundary condition \eqref{eq:E_dE/dn_BC} together with the condition that the two tangential components of $\bs{E}^\rmtot$ vanish on the boundary. These constitute the necessary and sufficient conditions to determine the scattered field $\bs{E}^\rmsc$ and also to ensure that $\bs{E}^\rmsc$ is divergence-free as required. 

Similarly, the magnetic field can be found by solving the surface integral equation corresponding to \eqref{eq:BIE_Ei} for the scattered $\bs{H}^\rmsc$ field with the boundary condition $\bs{n} \cdot \bs{H}^\rmtot = 0$ at the PEC surface. To apply the boundary condition on the tangential components of $\bs{E}^\rmtot$, we choose two orthogonal unit tangents $\bs{t}_1$ and $\bs{t}_2$ on $S$, and use Ampere's law to express the component of $\bs{E}^\rmtot$ parallel to say, $\bs{t}_1$, namely, $E_{t_1}^\rmtot = \bs{t}_1 \cdot \bs{E}^\rmtot = \bs{E}^\rmtot \cdot (\bs{t}_2 \times \bs{n})$, in terms of $\bs{H}^\rmtot$~\cite{Klaseboer2017}
\begin{subequations} \label{eq:E_tang_H}
  \begin{align} 
    E_{t_1}^\rmtot &= \bs{t}_2 \cdot \left(
                     \bs{n} \times \bs{E}^\rmtot \right)
                     = \frac{\rmi}{\omega \epsilon}
                     \left[ \bs{t}_2 \cdot \left(
                     \bs{n} \times \bs{\nabla} \times \bs{H}^\rmtot
                     \right) \right] \\
                   &= \frac{\rmi}{\omega \epsilon}
                     \left[ \bs{n} \cdot \left(
                     \bs{t}_2 \cdot \bs{\nabla}\right)
                     \bs{H}^\rmtot -  \bs{t}_2 \cdot (\bs{n} \cdot \bs{\nabla})
                     \bs{H}^\rmtot \right] = 0
  \end{align}
\end{subequations}
with a similar expression for $E_{t_2}^\rmtot $ obtained by interchanging subscripts 1 and 2 in \eqref{eq:E_tang_H}.

\section{Numerical Implementation}\label{sec:comp-advant-appl}
The boundary integral solution of~\eqref{eq:BIE_Ei} for components of the field is conceptually straightforward. For any scalar function $p(\bs r)$ that satisfies the Helmholtz equation, Green's second identity gives a surface integral relation between $p(\bs{r})$ and its normal derivative $\partial{p}/\partial{n}$ at points $\bs{r}$ and $\bs{r}_0$ on the boundary. But all singularities associated with Green's function $G(\bs{r},\bs{r}_0)$ can in fact be removed analytically to give~\cite{Klaseboer2012, Sun2015}
\begin{multline} 
  \label{eq:NS_BIE}
  \int_{S+S_{\infty}} \left[
    p(\bs{r}) - p(\bs{r}_0) g(\bs{r})
    - \frac{\partial p(\bs{r}_0)}{\partial n}  f(\bs{r}) \right]
  \frac{\partial G}{\partial n} \, \rmd S(\bs{r}) \\
  =  \int_{S+S_{\infty}} \left[
    \frac{\partial p(\bs{r})}{\partial n}
    - p(\bs{r}_0) \frac{\partial g(\bs{r})}{\partial n} 
    - \frac{\partial p(\bs{r}_0)}{\partial n} 
    \frac{\partial f(\bs{r})}{\partial n}
  \right] G \, \rmd S(\bs{r}), 
\end{multline}
where $S_\infty$ denotes an artificial surface at infinity.  The requirement on $f(\bs{r})$ and $g(\bs{r})$ is that they satisfy the Helmholtz equation and the following conditions at $\bs{r} = \bs{r}_0$ on the surface $S$: 
\begin{subequations}
  \label{eq:f_g_conditions}
  \begin{align}
    f(\bs{r}) = 0 \qquad \text{and} \qquad
    \bs{n} \cdot  \bs{\nabla} f(\bs{r}) = 1, \\
    g(\bs{r}) = 1 \qquad \text{and} \qquad
    \bs{n} \cdot \bs{\nabla} g(\bs{r}) = 0.
  \end{align}
\end{subequations}
Examples of simple choices of $f(\bs{r})$ and $g(\bs{r})$ can be found in~\cite{Klaseboer2012,Sun2015}. Thus, if $p$ (or $\partial{p}/\partial{n}$) is given, then~\eqref{eq:NS_BIE} can be solved for $\partial{p}/\partial{n}$ (or $p$) in a straightforward manner. This is because if $f(\bs{r})$ and $g(\bs{r})$ obey \eqref{eq:f_g_conditions}, then the terms that multiply $G$ and $\partial G/\partial n$ vanish at the same rate as the rate of divergence of $G$ or $\partial G/\partial n$ as $\bs{r} \rightarrow \bs{r}_0$. Consequently, both integrals in \eqref{eq:NS_BIE} have non-singular integrands, and thus can be evaluated accurately by quadrature, see~\cite{Klaseboer2012, Sun2015} for details. Note that the solid angle, $c_0$ at $\bs{r}_0$ has also been eliminated in \eqref{eq:NS_BIE}.

In our implementation, $p$ and $\partial p/\partial n$ are the unknowns to be solved for at the chosen Nyquist nodes on the surface. The surface shape is represented by quadratic surface elements anchored at these nodes and the variation of function values within these elements are found by quadratic interpolation from the nodal values~\cite{Klaseboer2017, Sun2017}. Since the integrals do not have divergent kernels, the integration can be evaluated using the standard Gauss quadrature. This facilitates the reduction of the number of degrees of freedom while increasing numerical precision. Also with high order surface elements, the surface geometries can be represented more faithfully than with planar elements. Finally, with the help of \eqref{eq:E_dE/dn_BC}, a matrix system can be constructed where the unknowns are $\bs{n} \cdot \bs{E}^\rmsc$ and the two tangential components of $\partial \bs{E}^\rmsc/\partial n$ at each node as detailed below.  

Given that the tangential components of the total electric field vanish on the PEC surface, we can express the scattered field in terms of its normal component and the tangential components of the incident field  $\bs{E}^\rminc$ as 
\begin{equation}
  \label{eq:E_nt}
  \bs{E}^\rmsc =  E_{n}^\rmsc \bs{n} - E_{t_1}^\rminc \bs{t}_1
  - E_{t_2}^\rminc  \bs{t}_2.
\end{equation}
Similarly, we can decompose the normal derivative of the scattered field on the surface of the scatterer in terms of its normal and tangential components to obtain
\begin{equation}\label{eq:dEdn_nt}
  \frac{\partial \bs{E}^\rmsc}{\partial n}
  = \left(\bs{n} \cdot \frac{\partial \bs{E}^\rmsc}{\partial n}\right) \bs{n}
  + \left(\bs{t}_1 \cdot \frac{\partial \bs{E}^\rmsc}{\partial n}\right)
  \bs{t}_1
  + \left( \bs{t}_{2} \cdot \frac{\partial\bs{E}^\rmsc}{\partial n}\right)
\bs{t}_2.
\end{equation}
Also, condition \eqref{eq:E_dE/dn_BC} can be written in terms of the scattered and incident fields as
\begin{equation}
  \label{eq:Esc_dE/dn_BC}
  \bs{n} \cdot \frac{\del \bs{E}^\rmsc}{\del n}
  = \left(\bs{n} \cdot \bs{E}^\rmsc\right) \kappa
  +\left(\bs{n} \cdot \bs{E}^\rminc \right)  \kappa
  -\bs{n} \cdot \frac{\del \bs{E}^\rminc}{\del n}.
\end{equation}
Introducing \eqref{eq:Esc_dE/dn_BC} into \eqref{eq:dEdn_nt}, then using the result together with \eqref{eq:E_nt} in the nonsingular surface integral equation \eqref{eq:NS_BIE} for each Cartesian component of the scattered field, we obtain the desired linear system.  Namely,
\begin{subequations}
  \label{eq:linear_system}
  \begin{equation}
    \label{eq:linear_system_part_a}
    \begin{bmatrix}
      n_{x} ({\cal{H}} - \kappa {\cal{G}}) & -t_{1x} {\cal{G}} & -t_{2x} {\cal{G}} \\
      n_{y} ({\cal{H}} - \kappa {\cal{G}}) & -t_{1y} {\cal{G}} & -t_{2y} {\cal{G}} \\
      n_{z} ({\cal{H}} - \kappa {\cal{G}}) & -t_{1z} {\cal{G}} & -t_{2z} {\cal{G}}
    \end{bmatrix}
    \begin{bmatrix}
      E_n^\rmsc \\ \bs{t}_{1} \cdot \frac{\partial \bs{E}^\rmsc}{\partial n} \\ \bs{t}_{2} \cdot \frac{\partial \bs{E}^\rmsc}{\partial n}
    \end{bmatrix}
    =
    \begin{bmatrix}
      {\cal{C}}_{x}\\ {\cal{C}}_{y} \\ {\cal{C}}_{z} 
    \end{bmatrix},
  \end{equation} 
  where
  \begin{equation}
    \label{eq:linear_system_part_b}
    {\cal{C}}_{\alpha} =  {\cal{H}} \left(t_{1\alpha}  {E}^{\text{inc}}_{t_{1}} + t_{2\alpha}  {E}^{\text{inc}}_{t_{2}} \right)  
    + n_{\alpha} {\cal{G}}\left(\kappa {E}^{\text{inc}}_{n} - \bs{n} \cdot \frac{\partial \bs{E}^\rminc}{\partial n} \right), \quad \alpha=x,y,z.
  \end{equation}
\end{subequations}
In \eqref{eq:linear_system}, ${\cal{G}}$ and ${\cal{H}}$ are the numerical matrix versions of \eqref{eq:NS_BIE} such that ${\cal{H}} \cdot p = {\cal{G}} \cdot \partial p/\partial n$, where $p$ and $\partial p/\partial n$ are now column vectors, see \ref{Appendix_GH}.
\begin{figure}[!t]
  \centering{}\includegraphics[width=0.7\textwidth]{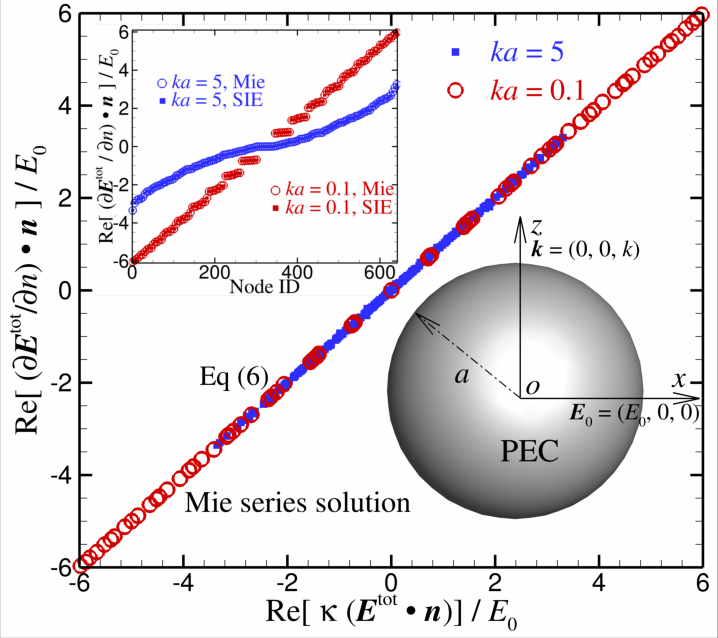}
  \caption{ \label{fig:Mie}
    A graphical confirmation of the boundary condition \eqref{eq:E_dE/dn_BC} is shown.  This boundary condition follows from enforcing $\bs{\nabla} \cdot \bs{E}^\rmtot = 0$ on the boundary and  relates $\bs{n} \cdot \bs{E}^\rmtot $ to $\bs{n} \cdot (\del \bs{E}^\rmtot/\del n)$. The $\bs{n} \cdot \bs{E}^\rmtot $ and $\bs{n} \cdot (\del \bs{E}^\rmtot/\del n)$ quantities are obtained from the Mie series solution with $ka =0.1$ (open red symbols) and $ka=5$ (solid blue symbols). Inset: Comparison of the field gradient $\bs{n} \cdot (\del \bs{E}^\rmtot/\del n) $ (ordered by numerical value) is shown at $642$ nodes connecting $320$ quadratic area elements that span the PEC surface obtained via our surface integral equation (SIE) method (solid symbols) and via the Mie series (open symbols).}
\end{figure}

\section{Illustrative Numerical Results}\label{sec:illustr-numer-result}
Consider scattering of the plane wave $\bs{E}^\rminc = E_0 \exp(\rmi k z)\bs{x}$ from a PEC sphere of radius $a$.  In \figurename~\ref{fig:Mie}, a graphical confirmation of the validity of the boundary condition \eqref{eq:E_dE/dn_BC} is shown, where each term in \eqref{eq:E_dE/dn_BC} is calculated using the analytical Mie series solution~\cite{Liou1977}.  The equality of the two sides of \eqref{eq:E_dE/dn_BC} for $ka = 0.1$ and $ka=5$, a 50-fold variation in $ka$, at various points on the sphere is clearly evident. Note that the sphere with a smaller radius (larger mean curvature, $\kappa = 2/a$) has larger range of magnitudes for the field gradient.

The accuracy of our surface integral equation (SIE) method to calculate the field gradient $\bs{n} \cdot (\del \bs{E}^{\text{tot}}/\del n) $ is demonstrated in the inset of \figurename~\ref{fig:Mie}. The relative difference between our SIE results and those from the Mie series is less than $1.6\%$ when $|\bs{n} \cdot (\del \bs{E}^{\text{tot}}/\del n)/E_0| \sim 1$. This field gradient is found directly in our formulation and does not require additional post-processing that might reduce the numerical accuracy.

\begin{figure}[!t]
  \centering{}\includegraphics[width=0.5\textwidth]{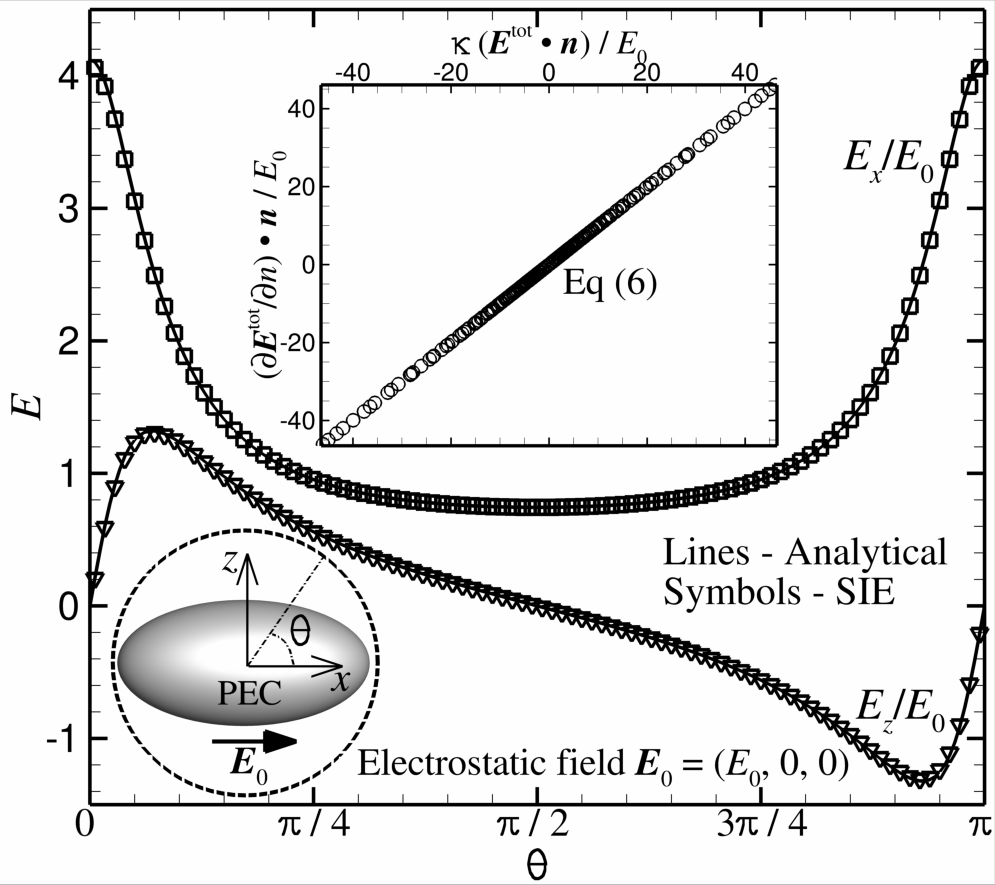} \caption{\label{fig:spheroid_k=0} 
    Comparisons of the electric field along the dotted circle of radius $1.05$ times the semi-major axis in the plane $y=0$ obtained from our SIE method at $k=0$ with $2562$ nodes and $1280$ quadratic elements (open symbols) and the analytical solution~\cite{Stratton1941} (lines).  Inset: Graphical confirmation of the boundary condition \eqref{eq:E_dE/dn_BC} based on the analytical solution~\cite{Stratton1941}.} 
\end{figure}

It is well-known that the surface current formulation of computational electromagnetics suffers from a numerical instability in the long wavelength (zero frequency) electrostatic limit due to the use of the charge-current continuity condition~\cite{Vico2016}. In contrast, the present field-only formulation does not suffer from this deficiency. To demonstrate this, we consider a PEC prolate spheroid of aspect ratio 2:1 embedded in an \emph{electrostatic} field ($k=0$) and polarized along the $x$-axis. In \figurename~\ref{fig:spheroid_k=0}, we compare the electric field at points on a circle of radius $1.05$ times the major semi-axis around the ellipsoid with the analytical solution~\cite{Stratton1941}. In the inset of \figurename~\ref{fig:spheroid_k=0}, we also confirm the boundary condition \eqref{eq:E_dE/dn_BC}.

Strictly speaking, our method requires a well-defined tangent plane, and thus it is only strictly valid for smooth scatterers.  Previously, we applied our \emph{singular} integral surface equations given by \eqref{eq:BIE_Ei} to a cube with sharp edges~\cite{Yuffa2019}.  Even though it is theoretically difficult to justify using \eqref{eq:BIE_Ei} on \emph{non-smooth} scatterers, we found that our \emph{singular} SIE performed better than the PMCHWT in terms of the condition number of the matrix system with respect to the mesh size~\cite{Yuffa2019}.  Armed with these encouraging results, we applied our \emph{nonsingular} formulation to a cube with \emph{rounded} edges, where the radius of the edge was chosen so that $r_{\text{edge}} \ll \lambda $.  The radar cross-sections (RCSs) obtained for our smoothed cube agreed well with the RCSs obtained for a cube with sharp edges for simple orientations considered in~\cite{Penno1989}.

\begin{figure}[!t]
\centering{}\includegraphics[width=0.75\textwidth]{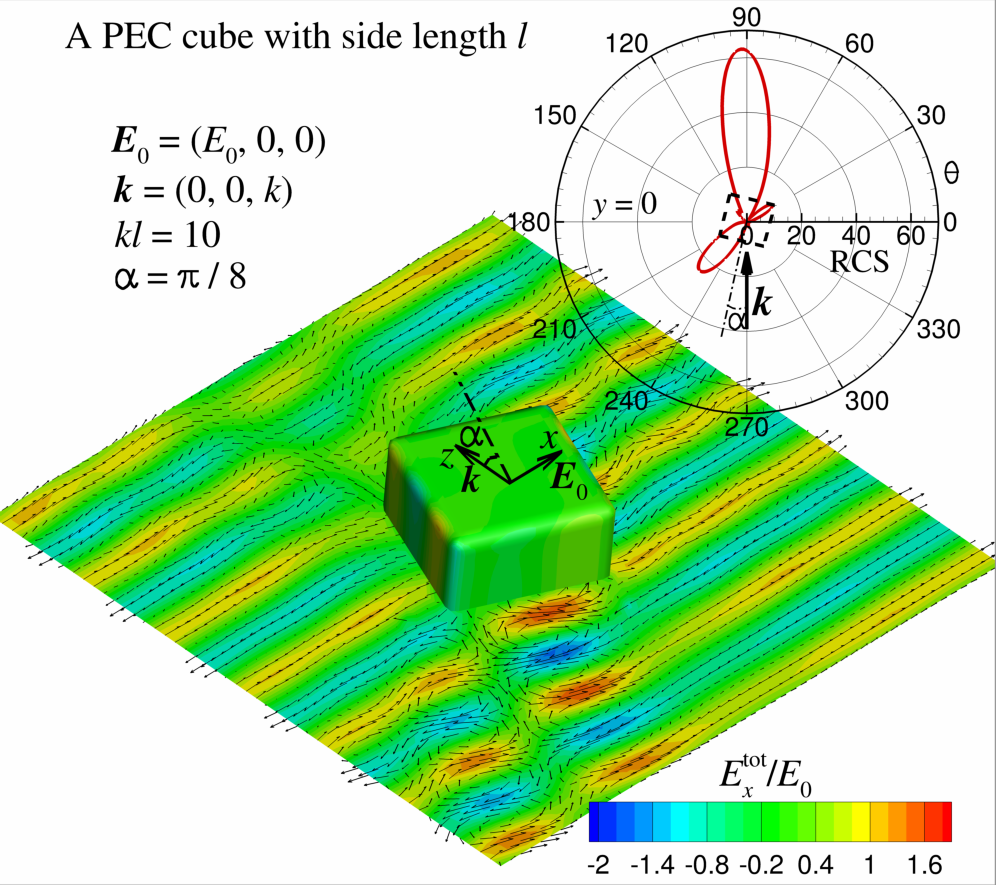} \caption{\label{fig:cube} 
The total electric field $\bs{E}^\rmtot$ and radar cross-section (inset) in the plane $y = 0$ around a PEC cuboid with rounded corners due to an incident electric field $\bs{E}^\rminc = E_0 \exp(\rmi kz) \bs{x}$. Results are obtained using $3482$ nodes and $1740$ quadratic elements.} 
\end{figure}

In \figurename~\ref{fig:cube}, we show a new example of the field in the plane $y = 0$ around a PEC cuboid with side $l$ and $kl = 10$ that is oriented at the angle $\alpha = \pi/8$ to the propagation direction of the incident plane wave. The center of the cuboid is located at the origin. The total field vectors are shown, and the magnitude of the $x$-component of the total field (the component parallel to the polarization of the incident field) is illustrated on a color scale. The radar cross-section on the $y = 0$ plane in the far field is shown in the inset of \figurename~\ref{fig:cube}.%

\section{Conclusions}\label{sec:conclusions}
The field-only formulation of computational electromagnetics developed here has a number of meritorious features:
\begin{enumerate}
    \item The formulation is conceptually simple, it only involves solving the Helmholtz equation for components of the field with the divergence-free constraint as a boundary condition.
    \item Physically important values of the field and its normal derivative at the surface are obtained directly without the need to work with intermediate quantities such as the surface current densities.
    \item The method does not involve hypersingular integrals~\cite{Chew2008_2} and is unaffected by the zero frequency catastrophe~\cite{Vico2016}.
    \item Technically, the solutions of scalar Helmholtz equations are implemented using a recently developed nonsingular method~\cite{Sun2015, Klaseboer2017, Sun2017} employing quadratic surface elements that affords higher precision with fewer degrees of freedom as well as being better able to represent the geometry of the surface than planar surface elements. 
    \item The reliance on only finding the solutions of scalar Helmholtz equations may also be advantageous in solving time-domain scattering problems using the inverse Fourier transform~\cite{Klaseboer2017_2}. 
\end{enumerate}

In short, the constructed framework is accurate, relatively easy to implement and works directly with the quantities of physical interest.


\renewcommand{\thesection}{Appendix A}
\section{Divergence-free Condition \label{Appendix_zero_div}}
\begin{figure}[!t]
  \centering
  \includegraphics[width=0.5\textwidth]{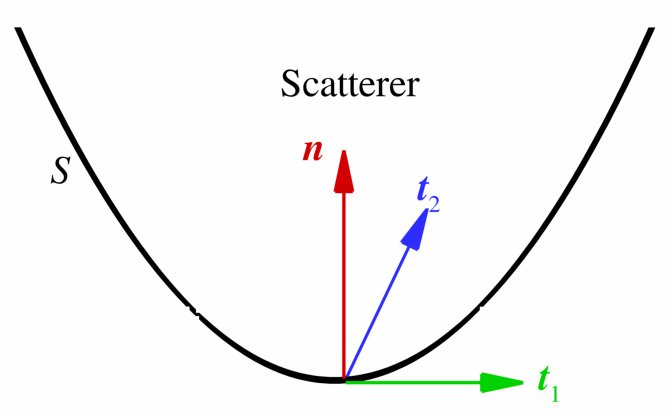}
  \caption{Sketch of the unit normal vector $\bs{n}$, the unit tangential vectors $\bs{t}_{1}$ with $\bs{t}_{1}\cdot \bs{n}=0$ and $\bs{t}_{2} = \bs{n} \times \bs{t}_{1}$ on the surface of the scatterer. \label{fig:nt1t2}}
\end{figure}

\setcounter{equation}{0}
\renewcommand{\theequation}{A\arabic{equation}}%
The divergence-free condition, $\bs{\nabla} \cdot \bs{E} = 0$  on $S$,  can be rewritten by decomposing $\bs{E}$ along the surface normal direction and the surface tangential plane, see \figurename~\ref{fig:nt1t2}.  The normal component and the two tangential components are defined by
\begin{equation}
  \label{eq:A_componets_defn}
  E_{n} \equiv \bs{n} \cdot \bs{E} , \quad 
  E_{t_1} \equiv \bs{t}_1 \cdot \bs{E}, \quad \text{and} \quad
  E_{t_2} \equiv \bs{t}_2 \cdot \bs{E},
\end{equation}
respectively.  Thus, by definition, on the surface $S$ we have
\begin{equation}
  \label{eq:A_del_defn}
  \bs{\nabla} \cdot \bs{E} \equiv
  \bs{\nabla} \cdot
  \left[ E_n \bs{n} + E_{t_1} \bs{t}_1 + E_{t_2}\bs{t}_2 \right].
\end{equation}
Expanding \eqref{eq:A_del_defn} yields
\begin{subequations}
  \label{eq:div_E_surf0}
\begin{equation}
  \label{eq:div_E_surf0_part_a}
  \bs{\nabla} \cdot \bs{E}= \left(\bs{n} \cdot \bs{\nabla}\right) E_n
  + E_n \bs{\nabla} \cdot \bs{n}
  + \left(\bs{t}_1 \cdot \bs{\nabla}\right) E_{t_1}
  + E_{t_2} \bs{\nabla} \cdot \bs{t}_2
  + \left(\bs{t}_2 \cdot \bs{\nabla}\right) E_{t_2}
  + E_{t_2} \bs{\nabla} \cdot \bs{t}_2,
\end{equation}
where
\begin{equation}
  \label{eq:div_E_surf0_part_b}
  \left(\bs{n} \cdot \bs{\nabla}\right) E_n =
  \bs{n} \cdot \left[ (\bs{n} \cdot \bs{\nabla}) \bs{E} \right]
  + \bs{E} \cdot \left[ (\bs{n} \cdot \bs{\nabla}) \bs{n} \right].
\end{equation}
\end{subequations}

In order to evaluate \eqref{eq:div_E_surf0}, we choose a local Cartesian coordinate system with the orthonormal basis $(\bs{x}, \bs{y}, \bs{z})$ located at the point $\bs{r}_{0}$ on the surface $S$. The local Cartesian coordinate system is aligned so that the normal at $\bs{r}_0$ is along $\bs{z}$ and the surface tangents are along $\bs{x}$ and $\bs{y}$. We will assume that the surface around $\bs{r}_0$ is locally quadratic and thus, the points $\bs{r} = (\xi_1, \xi_2, \zeta)$ that lie on the surface near $\bs{r}_0$ obey the relation
\begin{equation}
    \Phi(\xi_1, \xi_2, \zeta) \equiv \zeta -\textstyle \frac{1}{2} \kappa_1 \xi_1^2 - \textstyle \frac{1}{2} \kappa_2 \xi_2^2 = 0,
\end{equation}
where $\kappa_1$ and $\kappa_2$ are the principal curvatures at $\bs{r}_0$. The gradient operator in the local coordinates is 
\begin{equation}
    \bs{\nabla} \equiv \bs{x} \frac{\partial}{\partial \xi_1} + \bs{y} \frac{\partial}{\partial \xi_2} + \bs{z} \frac{\partial}{\partial \zeta},
\end{equation}
the unit normal at $\bs{r}$ is
\begin{equation}
    \bs{n} = \frac{\bs{\nabla} \Phi}{|\bs{\nabla} \Phi|} = \frac{( -\kappa_1 \xi_1,  - \kappa_2 \xi_2, 1)}{
    [1 + (\kappa_1 \xi_1)^2 + (\kappa_2 \xi_2)^2 ]^{1/2}},
\end{equation}
and the unit tangent vectors are
\begin{equation}
   \bs t_j = \frac{\partial \bs r/\partial \xi_j }{|\partial \bs r/\partial \xi_j|}, \quad \text{where} \quad j=1,2.
\end{equation}
Thus, at a point on $S$, i.e., $\bs r =(\xi_1, \xi_2,\zeta) = (\xi_1, \xi_2,  \textstyle \frac{1}{2} \kappa_1 \xi_1^2  + \textstyle \frac{1}{2} \kappa_2 \xi_2^2)$, we have
\begin{equation}\label{eq:surf_t1_t2}
    \bs{t}_{1} = \frac{ (1, 0, \kappa_1 \xi_1)  }{ [1 + (\kappa_1 \xi_1)^2]^{1/2} }  \;\;\text{and} \;\;
    \bs{t}_{2} = \frac{ (0, 1, \kappa_2 \xi_2)  }{ [1 + (\kappa_1 \xi_2)^2]^{1/2} }. 
\end{equation}
Furthermore, at $\bs{r} = \bs{r}_0 = (0, 0, 0)$, we also have the following identities:
\begin{subequations} \label{eq:geo_identities1}
\begin{eqnarray}
    \bs{\nabla} \cdot \bs{n} & = & -(\kappa_1 + \kappa_2) \equiv -\kappa, \label{eq:div_n_surf} \\
    \frac{\partial \bs{n}}{\partial n} & = & (\bs{n} \cdot \bs{\nabla}) \bs{n}  =  \bs{0}, \label{eq:n_grad_n_surf} \\
    \frac{\partial \bs{n}}{\partial t_1} & = & (\bs{t}_{1} \cdot \bs{\nabla}) \bs{n}  =  -\kappa_1 \bs{t}_{1}, \label{eq:t1_grad_n_surf} \\
    \frac{\partial \bs{n}}{\partial t_2} & = &(\bs{t}_{2} \cdot \bs{\nabla}) \bs{n}  =  -\kappa_2 \bs{t}_{2}, \label{eq:t2_grad_n_surf} 
\end{eqnarray}
\end{subequations}
\begin{subequations} \label{eq:geo_identities2}
\begin{eqnarray}
    \bs{\nabla} \cdot \bs{t}_{1} & = & 0, \label{eq:div_t1_surf} \\
    \frac{\partial \bs{t}_1}{\partial t_1} & = & (\bs{t}_{1} \cdot \bs{\nabla}) \bs{t}_{1}  =  \kappa_1 \bs{n}, \label{eq:t1_grad_t1_surf} \\
     \frac{\partial \bs{t}_1}{\partial t_2} & = & (\bs{t}_{2} \cdot \bs{\nabla}) \bs{t}_{1}  =  \bs{0}, \label{eq:t2_grad_t1_surf}
\end{eqnarray}
\end{subequations}
\begin{subequations} \label{eq:geo_identities3}
\begin{eqnarray}
    \bs{\nabla} \cdot \bs{t}_{2} & = & 0, \label{eq:div_t2_surf} \\
    \frac{\partial \bs{t}_2}{\partial t_1} & = & (\bs{t}_{1} \cdot \bs{\nabla}) \bs{t}_{2}  =  \bs{0}, \label{eq:t1_grad_t2_surf} \\
    \frac{\partial \bs{t}_2}{\partial t_2} & = & (\bs{t}_{2} \cdot \bs{\nabla}) \bs{t}_{2}  =  \kappa_2 \bs{n}, \label{eq:t2_grad_t2_surf}
\end{eqnarray}
\end{subequations}
where $\kappa=\kappa_1+\kappa_2$ is the mean curvature. Introducing identities \eqref{eq:geo_identities1}, \eqref{eq:geo_identities2}, \eqref{eq:geo_identities3} into \eqref{eq:div_E_surf0}, we obtain the desired result; namely,
\begin{equation}
  \label{eq:A_final_result}
  \bs{\nabla} \cdot \bs{E}  = \bs{n} \cdot \frac{\partial { \bs{E} }  }{ \partial{n} } - \kappa E_{n} + \frac{\partial { E_{t_{1}}  }} { \partial{t_{1}} } + \frac{\partial { E_{t_{2}}  }} { \partial{t_{2}} }=0.
\end{equation}%

\renewcommand{\thesection}{Appendix B}
\section{Numerical implementation of ${\cal{H}}$ or ${\cal{G}}$ \label{Appendix_GH}}
\setcounter{equation}{0}
\renewcommand{\theequation}{B\arabic{equation}}%
When the surface $S$ of the scatterer is discretised into $M$ (boundary) elements connected by $N$ nodes, the integrals in \eqref{eq:NS_BIE} can be evaluated by quadrature over each element which leads to the numerical matrix version of \eqref{eq:NS_BIE}, namely,
\begin{align}\label{eq:GH_NSBIM}
    {\cal{H}} \cdot p  = {\cal{G}} \cdot \frac{\partial{p}}{\partial{n}}.
\end{align}
In \eqref{eq:GH_NSBIM}, ${\cal{H}}$ represents influence matrix corresponding to the vector column $p$ at $N$ nodes, and ${\cal{G}}$ represents influence matrix corresponding to the vector column ${\partial{p}}/{\partial{n}}$ at $N$ nodes. 

\begin{figure}[!t]
  \centering
  \includegraphics[width=0.8\textwidth]{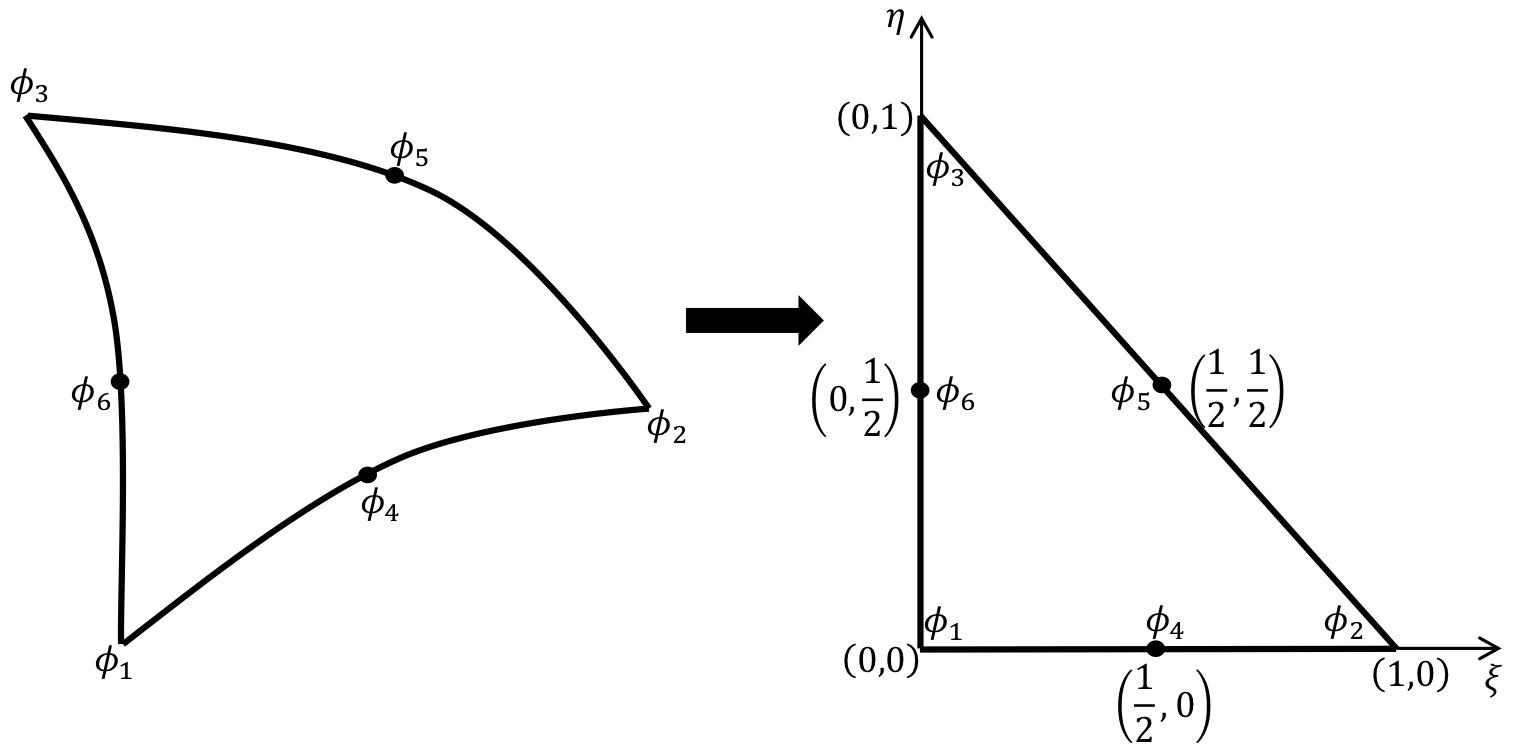}
  \caption{The quadratic interpolation scheme mapped onto a triangular surface element with the local surface variables ($\xi$, $\eta$), identical to those used in the finite element literature. \label{fig:triangleE}}
\end{figure}

For example, we can partition the surface $S$ into quadratic triangular area elements $S_{E}$ where each element is formed by three nodes on the vertices and three nodes on the edges. As shown in \figurename~\ref{fig:triangleE}, in terms of the local coordinates $(\xi, \eta)$ and $\nu \equiv 1 - \xi - \eta$, the coordinates of a point within each element and the function values at that point are expressed by quadratic interpolation from the values at the nodes using the standard quadratic interpolation function 
\begin{multline}
  \phi =  \nu(2\nu -1) \phi_{1} + \xi(2\xi -1) \phi_{2}+\eta(2\eta -1) \phi_{3}  
   +4 \nu\xi  \phi_{4} +4\xi \eta  \phi_{5} +4 \eta \nu  \phi_{6},
\end{multline}
where $\phi$ represents the position coordinates or function values within the element, and $\phi_{l=1,\ldots,6}$ are those values at all the nodes of the element. We employ standard mapping techniques as used in finite elements to transform each quadratic element into a triangular standard element on which Gaussian quadrature can be used (taking into account the Jacobian of transformation as well).

Consider the $i$th row of ${\cal{H}}$ or ${\cal{G}}$ that corresponds to the observation point $\bs{x}_0$ locating at node $i$ on $S$, for the items are off-diagonal that corresponds to the computation point $\bs{x}$ locating at node $j \neq i$ on $S$, we have
\begin{subequations} 
\begin{eqnarray}
    {\cal{H}}_{ij} & = & \sum_{l=1}^{m} \int_{S_{E_{l}}} C_{l} \frac{\partial{G}}{\partial{n}}  \text{ d} S_{E_{l}}, \\
    {\cal{G}}_{ij} & = & \sum_{l=1}^{m} \int_{S_{E_{l}}} C_{l} G  \text{ d} S_{E_{l}},
\end{eqnarray}
\end{subequations} 
where $m$ represents the number of elements that share node $j$ and $C_{l}$ corresponds to the weight for the node $j$ from the element $l$ based on the quadratic interpolation in (B2). For the diagonal entries, we have
\begin{subequations} 
\begin{eqnarray}
    {\cal{H}}_{ii} & = & \sum_{l=1}^{m} \int_{S_{E_{l}}} \left[C_{l} \frac{\partial{G}}{\partial{n}} - g(\bs{r}) \frac{\partial{G}}{\partial{n}} + \frac{\partial{g(\bs{r})}}{\partial{n}}  G \right]  \text{ d} S_{E_{i}} \nonumber \\
    && -  \sum_{l=1,l\neq i}^{M} \int_{S_{E_{l}}} g(\bs{r}) \frac{\partial{G}}{\partial{n}}  \text{ d} S_{E_{l}} +\sum_{l=1,l\neq i}^{M}  \int_{S_{E_{l}}}  \frac{\partial{g(\bs{r})}}{\partial{n}}  G \text{ d} S_{E_{l}}, \\
    {\cal{G}}_{ii} & = & \sum_{l=1}^{m} \int_{S_{E_{l}}} \left[ C_{l} G + f(\bs{r}) \frac{\partial{G}}{\partial{n}} -  \frac{\partial{f(\bs{r})}}{\partial{n}}  G \right]  \text{ d} S_{E_{l}} \nonumber \\
    && + \sum_{l=1,l\neq i}^{M} \int_{S_{E_{l}}} f(\bs{r}) \frac{\partial{G}}{\partial{n}}  \text{ d} S_{E_{l}} -\sum_{l=1,l\neq i}^{M}  \int_{S_{E_{l}}}  \frac{\partial{f(\bs{r})}}{\partial{n}}  G \text{ d} S_{E_{l}}
\end{eqnarray}
\end{subequations} 
where $M$ is the total number of the surface elements, $m$ is the number of elements that share the node $i$, and $C_{l}$ corresponds for the weight for the node $i$ from the element $l$ based on the quadratic interpolation in (B2).

It is worth emphasising again that the integrands in \eqref{eq:NS_BIE} are regular, so are those in (B3) and (B4). As such, all matrix entries of ${\cal{H}}$ or ${\cal{G}}$ can be obtained by using the standard quadrature, for example, a 12-point Gauss-Quadrature scheme to evaluate the integrals over each surface element.


\section*{Funding}
Australian Research Council (ARC) (DE150100169,  CE140100003, DP170100376).\\
This work was partially supported by U.S.~government, not protected by U.S.~copyright.


\section*{Disclosures}
The authors declare that there are no conflicts of interest related to this article.

 
\bibliography{refs.bib}

\end{document}